\documentclass[prb,longbibliography,showpacs,twocolumn,superscriptaddress,amsmath,amssymb,verbatim]{revtex4-1}

\usepackage{graphicx}
\usepackage{lmodern}
\usepackage{epstopdf}
\usepackage{dcolumn}
\usepackage{bm}
\usepackage{subfigure}
\usepackage{amsmath} 

\usepackage[pdftex,colorlinks=true,citecolor=blue,linkcolor=blue,urlcolor=blue,bookmarks=true]{hyperref}
\usepackage[sort&compress]{natbib}

\usepackage{color}
\usepackage{textcomp}
\definecolor{red}{rgb}{1,0,0}

\definecolor{blue}{rgb}{0,0,1}

\begin{document}
\preprint{APS}

\title{Coexisting spinons and magnons in frustrated zigzag spin-1/2 chain compound $\beta$-TeVO$_4$}
\author{M. Pregelj}
\email{matej.pregelj@ijs.si}
\affiliation{Jo\v{z}ef Stefan Institute, Jamova 39, 1000 Ljubljana, Slovenia}
\author{O. Zaharko}
\affiliation{Laboratory for Neutron Scattering and Imaging, PSI, CH-5232 Villigen, Switzerland}
\author{U. Stuhr}
\affiliation{Laboratory for Neutron Scattering and Imaging, PSI, CH-5232 Villigen, Switzerland}
\author{A. Zorko}
\affiliation{Jo\v{z}ef Stefan Institute, Jamova 39, 1000 Ljubljana, Slovenia}
\author{H. Berger}
\affiliation{Ecole polytechnique f\'{e}d\'{e}rale de Lausanne, CH-1015 Lausanne, Switzerland}
\author{A. Prokofiev}
\affiliation{Institute of Solid State Physics, Vienna University of Technology, Wiedner Hauptstrasse 8-10, 1040 Vienna, Austria}
\author{D. Ar\v{c}on}
\affiliation{Jo\v{z}ef Stefan Institute, Jamova 39, 1000 Ljubljana, Slovenia}
\affiliation{Faculty of Mathematics and Physics, University of Ljubljana, Jadranska c. 19, 1000 Ljubljana, Slovenia}
\date{\today}

\begin{abstract}

We investigate magnetic excitations in the frustrated zigzag spin-1/2 chain compound $\beta$-TeVO$_4$ by inelastic neutron scattering.
In the magnetically ordered ground state, the excitation spectrum exhibits coexisting magnon dispersion, characteristic of long-range magnetic order, and a spinon-like continuum that prevails above $\sim$2\,meV, indicating the dominance of intrachain interactions.
Combining linear-spin-wave-theory and pre-calculated spinon-continuum results, we reproduce the experimental spectrum.
Our analysis offers a minimal exchange-network model which determines dominant intrachain interactions, their anisotropies and weak interchain interactions.
The obtained parameters explain the magnetic ordering vector and spin excitations in the magnetic ground state.

\end{abstract}

\maketitle

\section{Introduction}

Frustrated magnetic systems are characterized by competing magnetic interactions, which allow for a number of different, yet energetically nearly degenerate, magnetic states.
As a result, already a subtle perturbation, e.g., a weak additional interaction or anisotropy, can completely change the magnetic ground state.
This leads to a vast variety of magnetic phases observed in such systems. \cite{lacroix2011introduction}
Prominent examples are highly-entangled disordered spin-liquid phases \cite{balents2010spin} on the one hand, and long-range-ordered spiral spin structures\cite{lacroix2011introduction} on the other hand, which besides the difference in the magnetic ordering also exhibit conceptually different excitations.
Namely, in classical long-range-ordered magnetic states, the elementary excitations are magnons (spin waves), i.e., collective spin-1 excitations of the magnetic order with precisely determined dispersion relations.\cite{des1962spin}
On the contrary, in spin liquids, typical excitations are spinons, i.e., deconfined fractional spin-1/2 excitations into which magnon fractionalize, that form a broad excitation continuum.\cite{lake2005soliton,zaliznyak2005quantum}

One of the simplest frustrated spin system is a zigzag spin-1/2 chain, where the nearest- and next-nearest-neighbor interactions compete.\cite{okunishi2003magnetic, hikihara2008vector, sudan2009emergent, hikihara2010magnetic}
The corresponding phase diagram encompasses spin-density-wave (SDW), vector-chiral (VC), as well as multipolar/spin-nematic phases, which are distinguished by specific one-dimensional correlations.
Yet, in most real systems, finite interchain interactions and anisotropies allow for spin correlations to expand in three dimensions and even lead to another magnetic ground state.\cite{furukawa2010chiral, schapers2013thermodynamic, du2016magnetic}
Such compounds display features of both one- and two-/three-dimensional spin systems. 
Namely, at low temperature and low energies a particular system may behave as a three-dimensionally ordered magnet,\cite{masuda2005spin, enderle2005quantum} while at higher energies, at elevated magnetic fields, or above the ordering temperature, a one-dimensional response may prevail.\cite{enderle2010two, orlova2017nuclear}
Examples of coexistence of three-dimensional magnon and one-dimensional spinon excitations have been reported for uniform spin chains,\cite{lake2005quantum, bera2017spinon} yet for zigzag chains such reports are scarce.\cite{enderle2005quantum, enderle2010two}

\begin{figure}[b]
\centering
\includegraphics[width=\columnwidth]{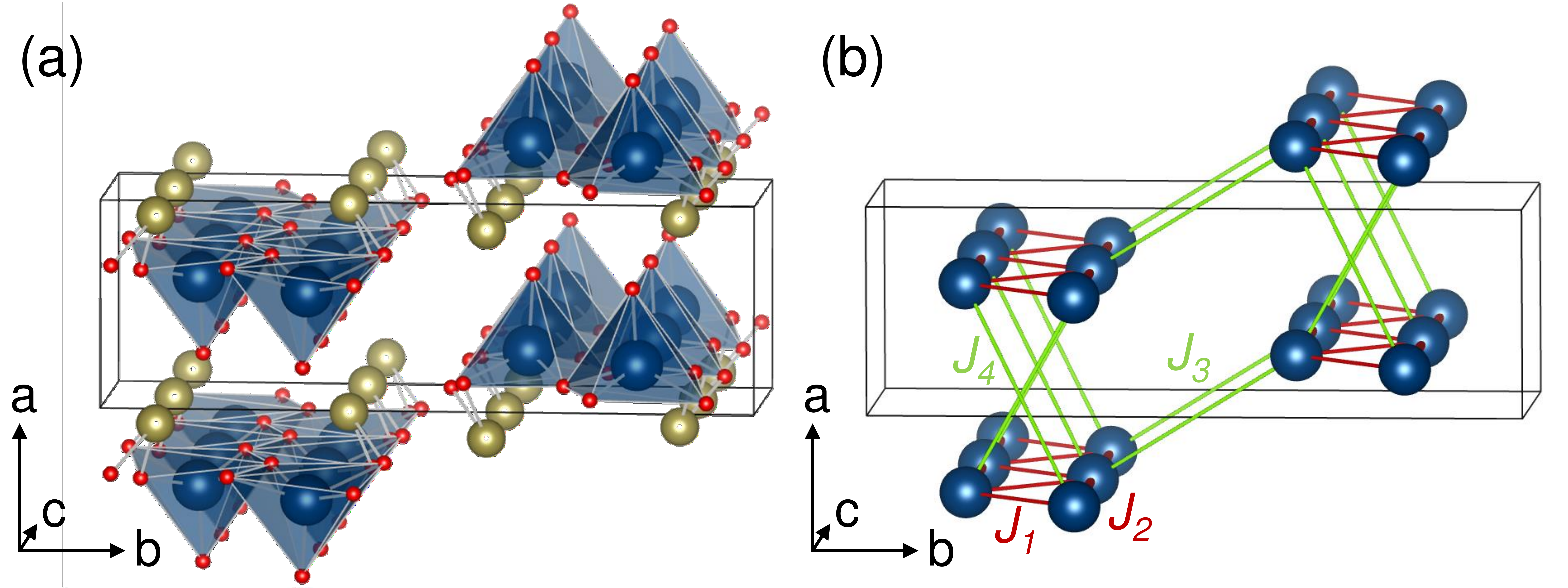}
\caption{(a) The crystal structure of $\beta$-TeVO$_4$. The coordination square pyramids of vanadium are shown in blue, oxygen ions are in red, and tellurium ions are light brown. The monoclinic unit cell ($Z$\,=\,4) is presented with a thin black line. (b) Magnetic exchange network in $\beta$-TeVO$_4$. The dominant intrachain couplings are shown in red ($J_1$ and $J_2$), while the weaker interchain couplings are shown in green ($J_3$ and $J_4$).}
\label{fig-network}
\end{figure}

Here we focus on a quasi-one-dimensional frustrated zigzag chain compound $\beta$-TeVO$_4$ [space group $P2_1/c$, V(4e), $a$\,=\,4.3919(1)\,\AA, $b$\,=\,13.5155(1)\,\AA, $c$\,=\,5.4632(1)\,\AA\, and $\beta$\,=\,90.779(1)$^\circ$, at $T$\,=\,10\,K].\cite{pregelj2015spin}
Spin chains are built of distorted corner-sharing VO$_5$ pyramids with magnetic V$^{4+}$ ($S$\,=\,1/2) ions that run along the crystalographic $c$ axis [Fig.\,\ref{fig-network}(a)]. \cite{meunier1973oxyde, savina2011magnetic}
The nearest-neighbor superexchange interaction $J_1$\,$\sim$\,$-$38\,K is ferromagnetic, while the next-nearest-neighbor interaction $J_2$\,$\sim$\,$-$0.8\,$J_1$ is antiferromagnetic, \cite{pregelj2015spin} thus imposing a strong magnetic frustration.
Additional interchain interactions were estimated to be an order of magnitude weaker and also frustrated.
These are responsible for a sequence of low-temperature magnetic transitions. 
At $T_{N1}$\,=\,4.65\,K the paramagnetic phase is replaced by an SDW phase with the incommensurate amplitude-modulated (ICAM) magnetic order defined by the magnetic wave vector {\bf k}\,$\approx$\,($-$0.20,\,0,\,0.41). \cite{pregelj2015spin, pregelj2016exchange}
The SDW phase is succeeded by an unconventional spin-stripe phase, which develops at $T_{N2}$\,=\,3.28\,K.
In this phase an additional weak ICAM magnetic component with {\bf k'}\,=\,({\bf k}+{\bf $\Delta$k}) and {\bf $\Delta$k}\,$\approx$\,($-$0.03,\,0,\,0.02) introduces a nanometer-sized stripe modulation of the spin structure.
Finally, at $T_{N3}$\,=\,2.28\,K, {\bf $\Delta$k} becomes zero and an elliptical-spiral magnetic order is established reflecting the dominance of the VC correlations at low temperatures.\cite{pregelj2016exchange}
Clearly, $\beta$-TeVO$_4$ can be rather well described as a frustrated zigzag spin-1/2 chain, yet the fascinating spin-stripe phase must emerge due to perturbative effects, i.e., probably due to weak interchain interactions and sizable exchange anisotropy ($\sim$20\%) of $J_1$ and $J_2$  that compete with fourth-order spin couping terms.\cite{pregelj2016exchange} 
A comprehensive description thus requires a detailed knowledge about the exchange network, which calls for an in-depth investigation of the magnon and spinon excitations.

In this study we investigate the magnetic excitation spectrum in the ground state of $\beta$-TeVO$_4$ using inelastic neutron scattering.
Our results reveal a coexistence of sharp magnon excitations and a broad spinon-like excitation continuum. 
The pronounced asymmetry of the continuum is in line with the fact that the $J_2/J_1$ ratio is close to $-$1.\cite{ren2012spinons}
Moreover, the modeling of the sharp dispersion by a linear spin-wave theory enabled us to estimate intrachain interactions and their anisotropies as well as to quantify the main weak interchain interactions.

\section{Experimental}

Inelastic neutron scattering experiments were performed using a sample consisting of five single crystals having a total mass of $\sim$0.8\,g.
The $a$ and $c$ axes defined the scattering plane.
The high-energy neutron scattering experiments, focused on excitations between 0.7 and 13\,meV, were performed on the thermal-neutron triple-axis spectrometer EIGER,\cite{stuhr2017thermal} at SINQ, at the Paul Scherrer Institute, Villigen, Switzerland.
The use of double focusing PG(002) monochromator and analyzer lead to the energy resolution at the elastic line of 0.7\,meV.
The final wave vector $k_f$\,=\,2.66\,\AA$^{-1}$ was filtered by a PG filter.
The low-energy experiments, focusing on excitations up to 3\,meV, were performed on the cold-neutron triple-axis spectrometer THALES, at the Institut Laue-Langevin, Grenoble, France.
In this experiment double focusing PG(002) monochromator, analyzer and a Be filter cooled to 77\,K (to remove the higher-order contamination) were used.
Here, $k_f$\,=\,1.55\,\AA$^{-1}$ and 1.3\,\AA$^{-1}$ setups lead to the resolution at the elastic line of $\sim$0.15\,meV and 0.1\,meV, respectively.
The temperature was in both cases controlled using standard ILL orange cryostats.
All measurements were performed at 1.5\,K.
In experiment performed at EIGER the sample crystals were co-aligned within $\delta\phi$\,$<$\,3$^\circ$, which is sufficient for the corresponding experimental resolution.
On the other hand, in the THALES experiment one crystal was shifted for  $\sim$12$^\circ$.
To avoid the spurious signal, the dispersion relations were measured at high reciprocal space values, i.e., the dispersion along $a^*$ near reflections with large $H$ and the dispersion along $c^*$ near reflections with large $L$.
A a result, the excitation spectra measured in the same part of the Brillouin zone at the two instruments differ only in resolution and signal to noise ratio, indicating that THALES data were not effected by misalignment of the sample.

\section{Results}

\subsection{Dispersion along the spin chains}

\begin{figure}[!]
\centering
\includegraphics[width=\columnwidth]{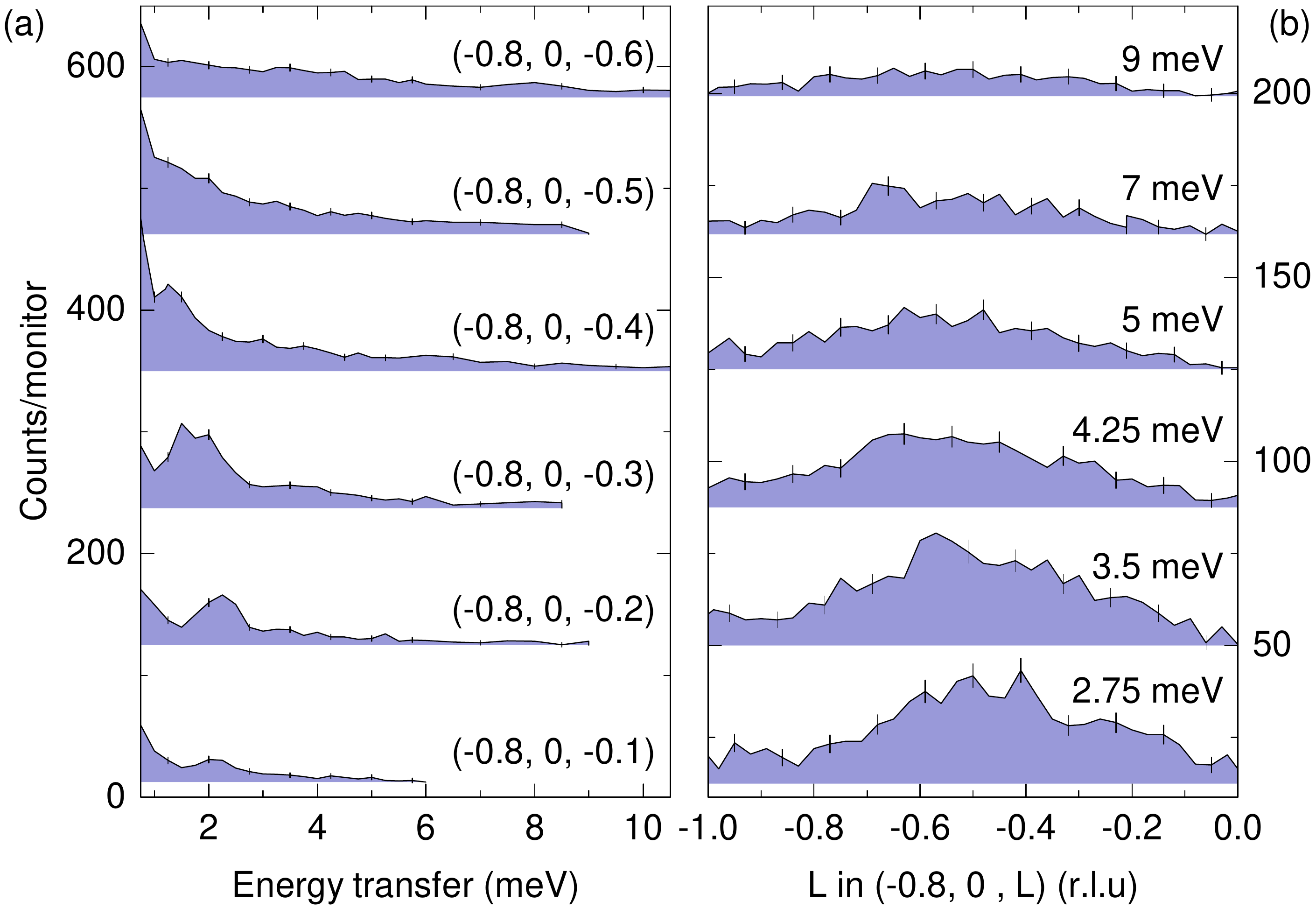}
\caption{(a) Energy scans at several positions in the reciprocal space measured at EIGER at 1.5\,K. Sharp features are magnon excitation modes, while broad background is due to the spinon continuum. (b) $L$ scans at ($-0.8,~0,~L$) at fixed energies. The broad feature implies the presence of the spinon continuum.}
\label{fig-Ecut-lcut}
\end{figure}

\begin{figure}[!]
\centering
\includegraphics[width=\columnwidth]{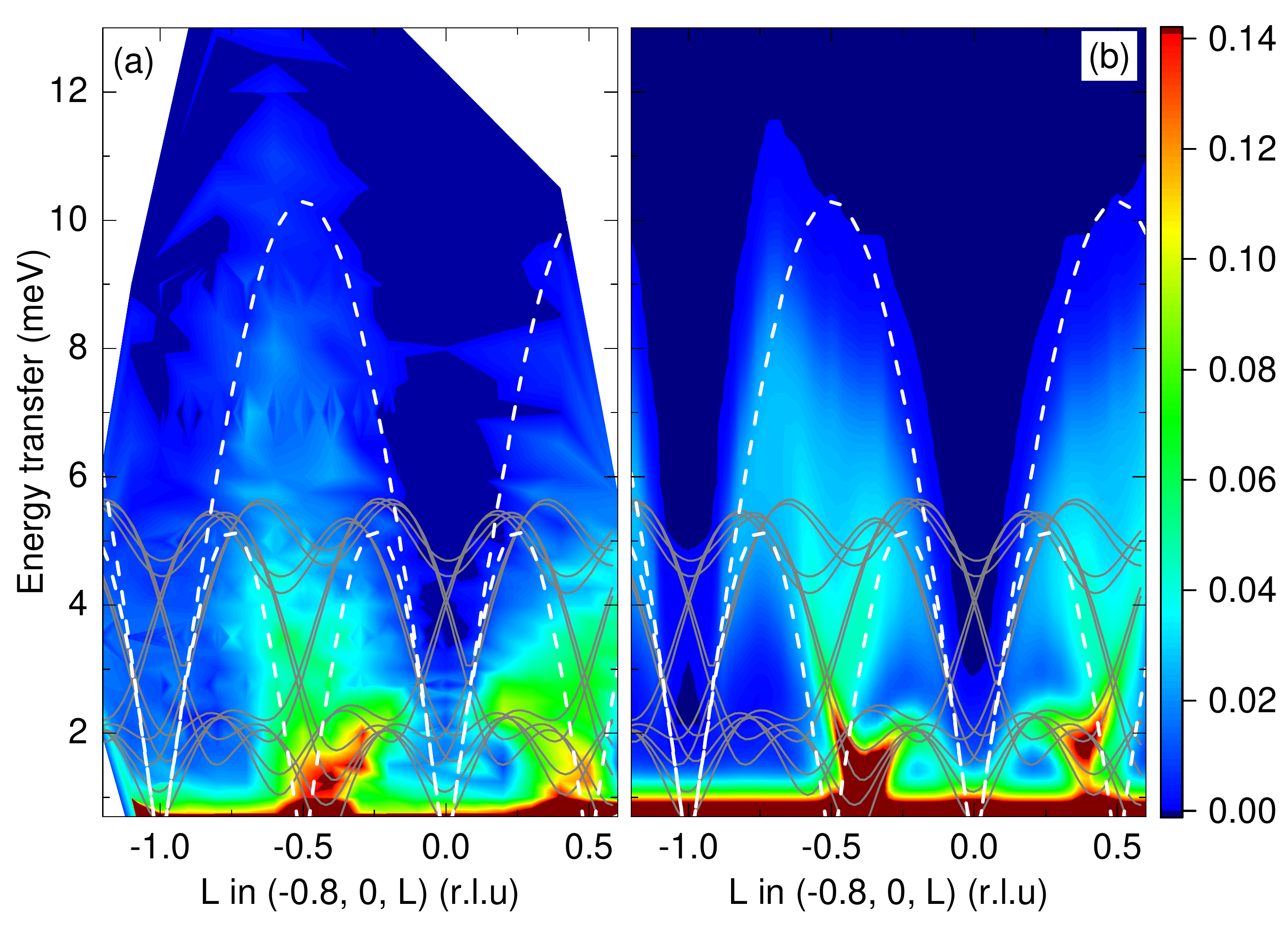}
\caption{(a) The energy map for ($-0.8,~0,~L$) cut in the reciprocal space measured at EIGER at 1.5\,K. (b) The calculated magnon dispersion for the model presented in text added to the spinon continuum calculated in Ref.\,\onlinecite{onishi2015magnetic}. In the calculated map, experimental resolution has been already taken into account. The solid lines in (a) and (b) represent the calculated spin-wave dispersion, while the dashed lines represent the upper and lower boundaries of a two-spinon continuum calculated for a uniform antiferromagnetic spin-1/2 chain, where the magnitude of the exchange $J$ equals $J_1$.}
\label{fig-ql}
\end{figure}

We first inspected the excitation spectrum along the spin chains (almost along $c^*$), which should be dictated mainly by the strong intrachain interactions.
We measured a series of energy scans at fixed positions in reciprocal space, crossing the magnetic reflections at ($-$0.8,~0,~$-$0.4) and ($-$0.8,~0,~0.6).
The excitation spectrum below 4\,meV exhibits several sharp features with the width comparable to the experimental resolution [Fig.\,\ref{fig-Ecut-lcut}(a)], which are therefore most likely associated with magnon excitations.\cite{des1962spin}
We note that $E$\,=\,4\,meV is comparable to the strength of $J_1$ and $J_2$, estimated to amount 3.3 and 2.6 meV, respectively. \cite{pregelj2015spin} 
In addition to these sharp features we find a broad excitation continuum extending across a wide region in energy and reciprocal space.
The existence of the latter is even more apparent when measuring the neutron-scattering intensity at fixed energies along $c^*$, i.e., changing the $L$ value [Fig.\,\ref{fig-Ecut-lcut}(b)].
The combined results are summarized in a map plot in Fig.\,\ref{fig-ql}(a) that reveals the coexistence of sharp features, extending up to $\sim$4\,meV, while a broad excitation continuum becomes pronounced above 2\,meV and extends up to $\sim$12\,meV.
The latter exhibits a dome-like shape, which reaches the maximum energy at $L$\,$\sim$\,0.5 r.l.u. and drops to zero energies at $L$\,$\sim$\,0 and 1 r.l.u., and is thus most probably associated with two-spinon excitations.\cite{lake2005soliton,zaliznyak2005quantum}

In the next step, we inspected the excitation spectrum along $c^*$ at lower energies and with higher resolution using THALES spectrometer at ILL.
However, we did not detect any additional excitation branches, which implies that weak additional interactions that could induce low-energy excitations must be perpendicular to the chains, as suggested by earlier studies. \cite{pregelj2015spin, saul2014density, weickert2016magnetic}

\subsection{Dispersion perpendicular to the spin chains}

\begin{figure}[!]
\centering
\includegraphics[width=\columnwidth]{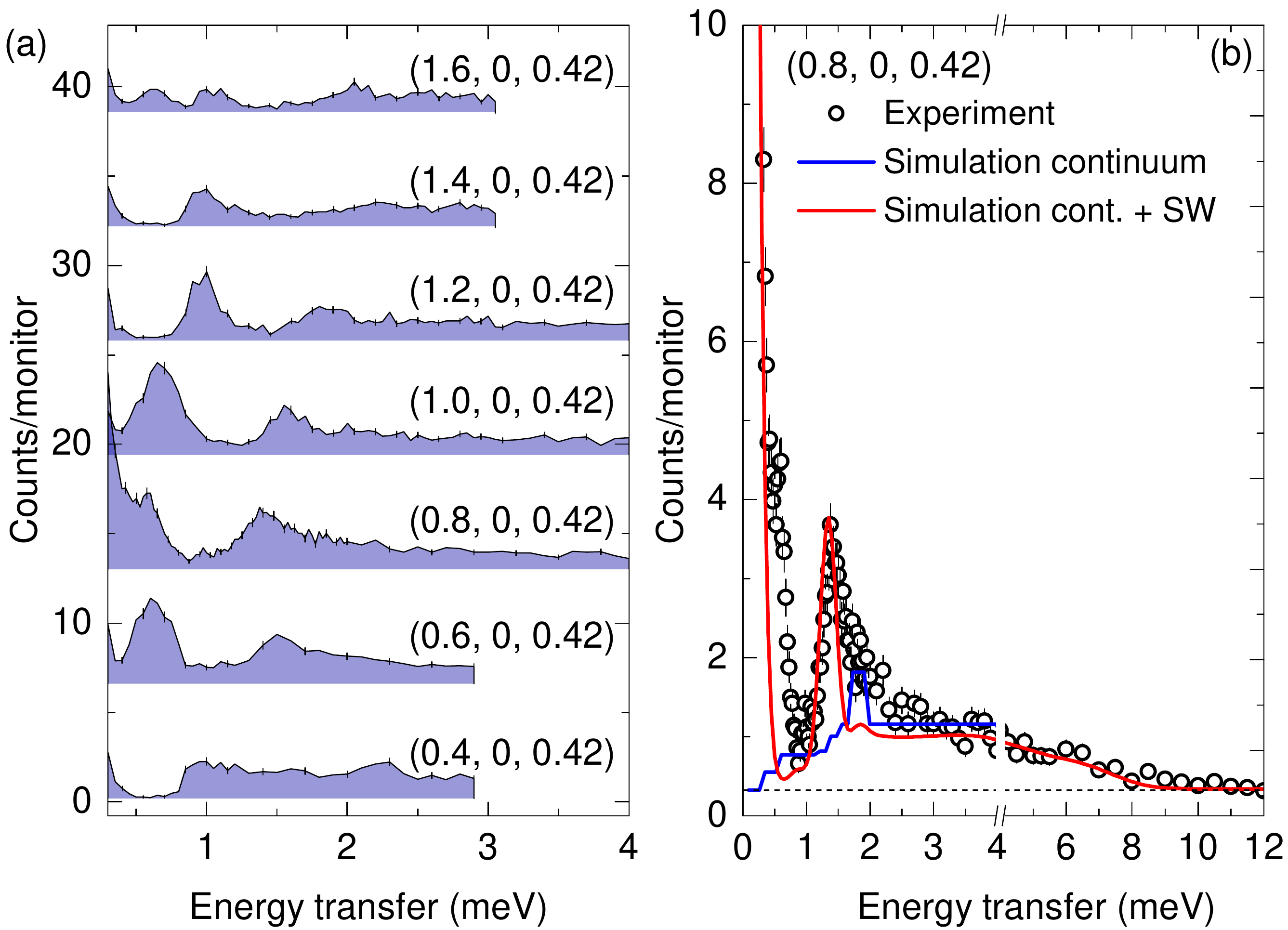}
\caption{(a) Energy scans at several positions in the reciprocal space measured at THALES at 1.5\,K. Sharp features are magnon excitation modes, while broad background is due to the spinon continuum. (b) Energy scan at the magnetic peak (0.8,~0,~0.42) measured at 1.5\,K (symbols) and simulation (line) for calculated magnon dispersion and spinon continuum calculated in Ref.\,\onlinecite{onishi2015magnetic}. In the calculated spectrum, experimental resolution has been already taken into the account.}
\label{fig-Ecut}
\end{figure}

\begin{figure}[!]
\centering
\includegraphics[width=\columnwidth]{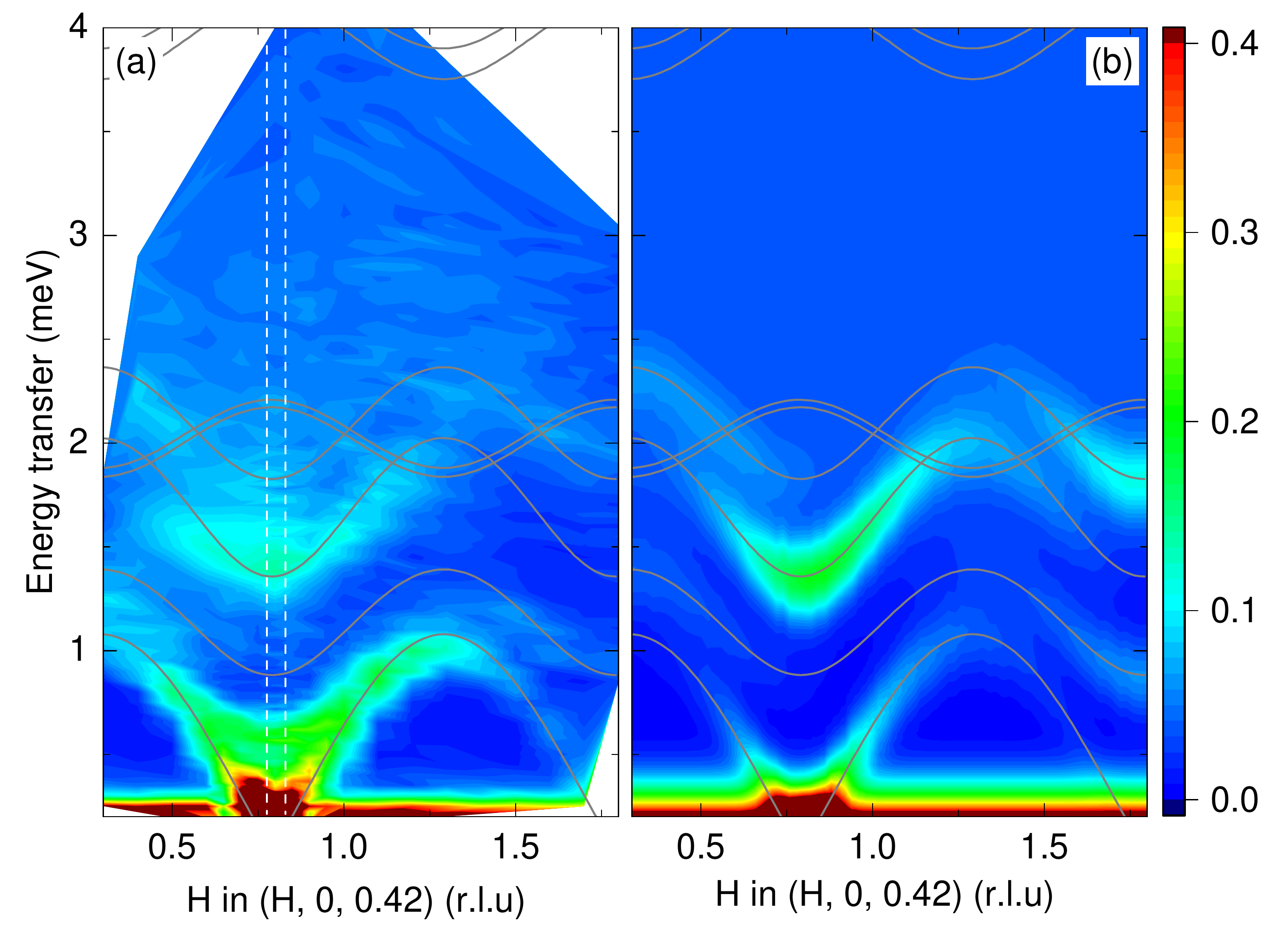}
\caption{(a) The energy map for ($H,~0,~0.42$) cut in the reciprocal space measured at THALES at 1.5\,K. The dashed vertical lines emphasize the scan shown in Fig.\,\ref{fig-Ecut}. (b) The calculated magnon dispersion for the model presented in text added to the spinon continuum calculated in Ref.\,\onlinecite{onishi2015magnetic}. In the calculated map, experimental resolution has been already taken into account. The solid lines in (a) and (b) represent the calculated spin-wave dispersion.}
\label{fig-qh}
\end{figure}

Next, we focused on the dispersion perpendicular to the chains, which should be significantly influenced by the weak interchain interactions.
In contrast to the measurements performed along $c^*$, we were unable to resolve any significant change in the excitation spectrum along $a^*$, when measuring with the high-energy setup at the EIGER spectrometer.
Hence, we performed additional high-resolution measurements at the THALES spectrometer at ILL. 
This allowed us to inspect excitations with energies above $\sim$0.3\,meV, as this is the point where the intensity of the elastic contribution falls below the background level.
We performed energy scans at fixed positions in reciprocal space along $a^*$, crossing the (0.8,~0,~0.4) magnetic reflection.
The excitation spectrum exhibits several sharp features with the width comparable to the experimental resolution (Fig.\,\ref{fig-Ecut}). 
In addition, we observed the broad excitation continuum above $\sim$2\,meV, which appears almost independent of $H$.
All results are summarized in the map plot in Fig.\,\ref{fig-qh}(a).
Obviously, sharp features are here better resolved than in the measurements along $c^*$, as for this orientation they appear at lower energies and are thus clearly separated from (they appear below) the continuum. 

%

\section{Discussion}

\subsection{Spinon continuum}

Our measurements reveal a broad excitation continuum that is characteristic of fractional spinon excitations, which are inherent to spin-1/2-chain systems.\cite{lake2005soliton,zaliznyak2005quantum}
The main contribution to the spinon continuum is due to two-spinon excitations, which for a uniform antiferromagnetic spin-1/2 chain occur between $E_{\text{low}}$\,=\,$\pi/2$\,$J|$sin($k$)$|$ and $E_{\text{high}}$\,=\,$\pi$\,$J|$sin($k/2$)$|$, where $J$ is the antiferromagnetic exchange constant.\cite{muller1981quantum, caux2006four, lake2005soliton, zaliznyak2005quantum}
The resulting 
dynamical structure factor -- the quantity measured by inelastic neutron scattering -- diverges at the lower boundary and decreases with increasing energy, $E$, as given by expression 1/($E^2$\,$-$\,$E_{\text{low}}^2(k)$)$^{1/2}$.\cite{muller1981quantum}
Assuming that the magnitude of $J$ equals $J_1$, we plot the corresponding two-spinon-continuum boundaries (Fig.\,\ref{fig-ql}).
Clearly, the measured continuum exists already below the lower theoretical boundary $E_{\text{low}}$, highlighting the discrepancy between $\beta$-TeVO$_4$ and the uniform spin-1/2 chain model.
In fact, our results are more similar to the response expected in frustrated zigzag spin-1/2 chain with ferromagnetic $J_1$, where the lower boundary is significantly lowered and becomes asymmetric with increasing next-nearest neighboring interactions.\cite{ren2012spinons, onishi2015magnetic}
Moreover, asymmetry and lowering of the spinon continuum has been observed in LiCuVO$_4$,\cite{enderle2010two} which is one of the most studied frustrated ferromagnetic zigzag spin-1/2 chain compounds.
However, in contrast to the latter, our data do not show significant scattering above the two-spinon continuum boundary $E_{\text{high}}$ (Fig.\,\ref{fig-ql}), suggesting that, if present, the contribution of four-spinon excitation in $\beta$-TeVO$_4$ is far less pronounced.

\subsection{Spin-wave modeling}

The acquired data provide detailed information about the interactions that constitute the exchange network in $\beta$-TeVO$_4$.
Previous experimental studies offered a good estimate of the main exchange interactions, $J_1$ and $J_2$, \cite{pregelj2015spin} an approximate estimation of their anisotropies, \cite{pregelj2016exchange} and a rough assessment of the interchain interactions. \cite{pregelj2015spin}
On the other hand, density-functional-theory calculations, \cite{saul2014density,weickert2016magnetic} investigating the exchange interactions up to the eight-shortest vanadium-vanadium distances, offered a quantitative estimate for all interactions included in calculations.
Since these calculations comply with previous experimental results,\cite{pregelj2015spin,pregelj2016exchange} the derived exchange parameters ensure a good foundation to describe present inelastic-neutron-scattering results as well as the modulation of the magnetic order perpendicular to spin chains.

In order to refine the exchange-network parameters we first inspect the origin of the magnetic modulation reflecting in the magnetic wave vector.
The two intrachain interactions $J_1$ and $J_2$ determine the magnetic modulation along the chains, \cite{pregelj2015spin} i.e., along $c^*$, while the modulation perpendicular to the chains (either along $a^*$ or $b^*$) depend on the weak interchain interactions.
In the reciprocal lattice units, the modulation period along $a^*$ is almost exactly half of the modulation period along $c^*$.
This can be explained already by a single exchange interaction that couples vanadium ions in the consecutive sites along the chain that are located in the neighboring chains, e.g., connecting sixth- or seventh-nearest-neighbor vanadium ions (the former being denoted as $J_4$ in Fig.\,\ref{fig-network}). 
On the other hand, the modulation along $b^*$ coincides with the crystal structure, indicating that the sequence of the exchange interactions between the two equivalent magnetic ions along the $b$ axis favors parallel alignment.
Hence, such sequence must involve an even number of antiferromagnetic interactions.

To describe the observed magnetic modulation one clearly has to take into account interchain interactions.
To avoid overparametrization, our goal is to find a minimal set of exchange parameters. 
Furthermore, as the magnetic wave vector in $\beta$-TeVO$_4$ is incommensurate along $a^*$ and $c^*$, we had to limit our analysis to coplanar magnetic structures with evenly sized magnetic moments to make calculations manageable.
Using the SpinW library \cite{toth2015linear} for numerically simulating magnetic structures and excitations based on the linear spin-wave theory, we thus tested different exchange combinations, involving interactions up to the twelfth-shortest distance between vanadium ions.
Surprisingly, we find that besides the intrachain interactions, $J_1$ and $J_2$, already two interchain interactions, $J_3$ and $J_4$ (Fig.\,\ref{fig-network}), corresponding to the fifth- and sixth-shortest distances between vanadium ions, \cite{saul2014density} are sufficient to reproduce the magnetic wave vector.
In particular, $J_3$ has to be antiferromagnetic and $J_4$ has to be ferromagnetic, which nicely complies with the density-functional-theory calculations. \cite{saul2014density, weickert2016magnetic}
We note that other compatible exchange combinations were dismissed, because they involved more interactions.

Next, we calculate spin-wave (magnon) dispersion relations based on the linear spin-wave theory, employing SpinW library, \cite{toth2015linear} and compare them with the inelastic neutron scattering results.
We consider exchange interactions $J_i$ ($i$\,=\,1-4), with $J_1$ and $J_2$ having a sizable exchange anisotropy along $b$, $\delta_i^b$ ($i$\,=\,1,2), i.e., imposing either easy-plane or easy-axis anisotropy that is slightly distorted by a small additional anisotropy along $c$, $\delta_i^c$  ($i$\,=\,1,2), as implied by the recent anisotropy study,\cite{pregelj2016exchange} while $J_3$ and $J_4$ were assumed to be completely isotropic.
The corresponding Hamiltonian thus has the form
\begin{align}
\begin{split}
\label{Hamiltonian}
H = J_1\sum_{n,j}({\bf S}_{n,j}\cdot{\bf S}_{n,j+1}+\delta^b_1 S^b_{n,j}S^b_{n,j+1}+\delta^c_1 S^c_{n,j}S^c_{n,j+1})\\
+ J_2\sum_{n,j}({\bf S}_{n,j}\cdot{\bf S}_{n,j+2}+\delta^b_2 S^b_{n,j}S^b_{n,j+2}+\delta^c_2 S^c_{n,j}S^c_{n,j+2})\\
+J_3\sum_{\langle n, m \rangle,j}{\bf S}_{n,j}\cdot{\bf S}_{m,j-1}
+J_4\sum_{\langle n, m' \rangle,j}{\bf S}_{n,j}\cdot{\bf S}_{m',j+1},
\end{split}
\end{align}
where $n$ labels the chains, $j$ the position of the spin along the chain, while $m$ and $m'$ denotes neighboring chains along the $b$ and $a$ axis, respectively.
To make calculations manageable, we assume that the size of all V$^{4+}$ magnetic moments is fixed to 1\,$\mu_B$, where $\mu_B$ is the Bohr magneton, and that the magnetic structure is coplanar. 
Finally, we consider also the presence of the spinon continuum, which appears to dominate the excitation spectrum above 2\,meV.
In particular, we combine our spin-wave calculations with the spinon contribution calculated previously by density-matrix-renormalization-group (DMRG) method for the $J_1$\,=\,$-$$J_2$ spin-1/2 chain\cite{onishi2015magnetic} and obtain the magnetic structure factor as a sum of two contributions
\begin{align}
\begin{split}
\label{SWplusCONT}
S(E,{\bf Q}) =~  &f(E)S_{\text{sw}}(E,{\bf Q}) + \\
& [1-f(E)]S_{\text{cont}}(E,{\bf Q}),
\end{split}
\end{align}
where $S_{\text{sw/cont}}(E,{\bf Q})$ is spin-wave/spinon-continuum contribution, $f(E)$\,=\, 1/($\exp[(E\,-\,E_c)/\delta E]+1)$, $E_c$ denotes the energy of the transition between spin-wave and spinon type of excitations that has the width of $\delta E$.
The best match with the experiment was obtained by adjusting $E_c$ and $\delta E$, while the experimental resolution was approximated by convolving the calculated energy spectra with Gaussian function with the width determined from the elastic incoherent vanadium scattering. 
The results showing the best agreement with the experiment are presented in Fig.\,\ref{fig-ql}(b) and Fig.\,\ref{fig-qh}(b), yielding {\bf k}$_{\text{IC}}$\,=\,($-$0.208,~0,~0.419) and magnetic moments lying in the $ac$ plane.
The corresponding parameters are $J_1$\,=\,$-$38\,K, $J_2$\,=\,38\,K, $J_3$\,=\,3\,K, $J_4$\,=\,$-$1.9\,K, $\delta_1^b$\,=\,0.106, $\delta_2^b$\,=\,$-$0.126, $\delta_1^c$\,=\,0.01, and $\delta_2^c$\,=\,0.01.
We evaluate the reliability of the derived parameters by inspecting their sensitivity to the changes of the magnetic-ordering plane, which yields the uncertainty of $\sim$5\% for $J_i$ ($i$\,=\,1-4), $\sim$20\% for $\delta_i^b$ ($i$\,=\,1,2), and $\sim$50\% for $\delta_i^c$ ($i$\,=\,1,2).
On the other hand, $E_c$ was found to roughly scale with the square root of the energy resolution, yielding $E_c$\,=\,1.3 and 2.2\,meV for THALES and EIGER instruments, respectively, whereas $\delta E$\,=\,0.5\,meV was the same.

The derived parameters are in good agreement with previous studies, \cite{pregelj2015spin, pregelj2016exchange, saul2014density,weickert2016magnetic} but also reveal a few interesting new points.
Namely, both large exchange anisotropies, $\delta^b_1$ and $\delta^b_2$, yield ferromagnetic contributions, i.e., the former increases the ferromagnetic $J_1$, while the latter reduces the antiferromagnetic $J_2$.
This suggests that they have similar origins, which could be associated with hopping of electrons between the ground states and excited states of vanadium ions in combination with the spin-orbit coupling.\cite{von2002anisotropic, eremin2006unconventional} 
However, since $J_1$ and $J_2$ have different signs, $\delta^b_1$ and $\delta^b_2$ impose different types of anisotropy, i.e., the former imposes an easy-axis anisotropy, while the latter imposes an easy-plane one.
As a result, the effective ratio $(J_1/J_2)^i$, i.e., $J_1(1+\delta^i_1)/J_2(1+\delta^i_2)$, for the $b$ spin component ($i$\,=\,$b$) differs by $\sim$20\% compared to the same ratio for the $a$ and $c$ spin components ($i$\,=\,$a,c$).
This complies with the estimate derived in Ref.\,\onlinecite{pregelj2016exchange} and thus corroborates the scenario that the exchange anisotropy imposes two distinct modulations in the spin-stripe phase, i.e., for the $a$-$c$ and for the $b$ spin component.
In addition, the weak anisotropy $\delta^c$ reduces the energy difference between magnetic moments lying in the $ab$ and $bc$ planes, which is in line with the perpendicular orientations of the cycloids in the adjacent chains observed in the ground state phase.\cite{pregelj2016exchange}
Moreover, our analysis evaluates the interchain couplings that induce long-range magnetic ordering and are essential for the formation of the intriguing spin-stripe phase.

Finally, we note that the derived parameters may deviate from the exact solution of (\ref{Hamiltonian}) due to approximations undertaken in the data analysis, e.g., coplanar magnetic structure and same-sized magnetic moments.
In addition, the magnetic ground-state minimization and linear-spin-wave-theory approach used in SpinW library are optimized for large, i.e., classical, spins and three-dimensional magnetic lattices.
The fact that the agreement between the experiment and calculations is not perfect (Figs.\,\ref{fig-ql} and \ref{fig-qh}) thus suggests that there could be other contributions to the spin Hamiltonian that were not identified.
Nevertheless, the overall agreement between the experimental observations and theoretical modeling suggests that despite its limitation the proposed exchange model is not far from the actual situation and thus represents a promising starting point for more involved theoretical studies.

\section{Conclusion}

In conclusion, our low-temperature inelastic neutron scattering results indicate the coexistence of magnon and spinon excitations in $\beta$-TeVO$_4$, which has been so far rarely observed in frustrated zigzag spin-1/2 chain compounds.
Employing the linear spin-wave theory we were able to model the observed dispersion and determined the main exchange interactions and anisotropies.
Our results comply with the preceding studies of $\beta$-TeVO$_4$ and thus support the hypothesis that the exchange anisotropy and interchain interactions are most likely responsible for the establishment of the peculiar spin-stripe phase.\cite{pregelj2015spin, pregelj2016exchange}
The derived exchange parameters thus open a way for detailed theoretical studies of the corresponding spin-stripe-formation mechanism that is still puzzling.

\begin{acknowledgments}
We acknowledge the financial support of the Slovenian Research Agency (projects Z1-5443 and program No. P1-0125) and the Swiss National Science Foundation (project SCOPES IZ73Z0\_152734/1).
The neutron diffraction experiments were performed at the Swiss spallation neutron source SINQ, at the Paul Scherrer Institute, Villigen, Switzerland, and at the reactor of the Institute Laue-Langevin, Grenoble, France.
We are grateful to M. Enderle for fruitful discussion and local support at ILL.
\end{acknowledgments}

\end{document}